# GREGoR: Accelerating Genomics for Rare Diseases

Moez Dawood[1,2,3], Ben Heavner[4], Marsha M. Wheeler[4], Rachel A. Ungar[5,6,7], Jonathan LoTempio[8], Laurens Wiel[5,6,9], Seth Berger[10,11,12], Jonathan A. Bernstein[13], Jessica X. Chong[14,15], Emmanuèle C. Délot[8], Evan E. Eichler[15,16,17], Richard A. Gibbs[1,2], James R. Lupski[1,2,18], Ali Shojaie[4], Michael E. Talkowski[19,20,21,22,23], Alex H. Wagner[24,25,26], Chia-Lin Wei[16], Christopher Wellington[27], Matthew T. Wheeler[9], GREGoR Partner Members, Claudia M. B. Carvalho[28], Casey A. Gifford[5,13,29,30], Susanne May[4], Danny E. Miller[15,16,31,32], Heidi L. Rehm[19,20], Fritz J. Sedlazeck[1,33], Eric Vilain[8], Anne O'Donnell-Luria[19,20,34], Jennifer E. Posey[2], Lisa H. Chadwick[35], Michael J. Bamshad[14,15,36], Stephen B. Montgomery[5,6,37], Genomics Research to Elucidate the Genetics of Rare Diseases (GREGoR) Consortium

[1]Human Genome Sequencing Center, Baylor College of Medicine, Houston, TX, USA.
[2]Department of Molecular and Human Genetics, Baylor College of Medicine, Houston, TX, USA.
[3]Medical Scientist Training Program, Baylor College of Medicine, Houston, TX, USA.
[4]Department of Biostatistics, University of Washington, Seattle, WA, USA.
[5]Department of Genetics, School of Medicine, Stanford University, Stanford, CA, USA.
[6]Department of Pathology, School of Medicine, Stanford University, Stanford, CA, USA.
[7]Stanford Center for Biomedical Ethics, School of Medicine, Stanford University, Stanford, CA, USA.
[8]Institute for Clinical and Translational Science, University of California, Irvine, CA, USA.
[9]Division of Cardiovascular Medicine, School of Medicine, Stanford University, Stanford, CA, USA.
[10]Division of Genetics and Metabolism, Children's National Rare Disease Institute, Washington, DC, USA.
[11]Center for Genetic Medicine Research, Children's National Rare Disease Institute, Washington, DC, USA.
[12]Department of Genomics and Precision Medicine, George Washington University, Washington, DC, USA.
[13]Department of Pediatrics, School of Medicine, Stanford University, Stanford, CA, USA.
[14]Department of Pediatrics, Dvision of Genetic Medicine, University of Washington, Seattle, WA, USA.
[15]Brotman Baty Institute for Precision Medicine, University of Washington, Seattle, WA, USA.
[16]Department of Genome Sciences, University of Washington, Seattle, WA, USA.
[17]Howard Hughes Medical Institute, University of Washington, Seattle, WA, USA.
[18]Department of Pediatrics, Baylor College of Medicine, Houston, TX, USA.
[19]Center for Genomic Medicine, Massachusetts General Hospital, Boston, MA, USA.
[20]Program in Medical and Population Genetics, Broad Institute of MIT and Harvard, Boston, MA, USA.
[21]Department of Neurology, Massachusetts General Hospital and Harvard Medical School, Boston, MA, USA.
[22]Stanley Center for Psychiatric Research, Broad Institute of MIT and Harvard, Cambridge, MA, USA.
[23]Program in Bioinformatics and Integrative Genomics, Harvard Medical School, Boston, MA, USA.
[24]Steve and Cindy Rasmussen Institute for Genomic Medicine, Nationwide Children's Hospital, Columbus, OH, USA.
[25]Department of Pediatrics, The Ohio State University College of Medicine, Columbus, OH, USA.
[26]Department of Biomedical Informatics, The Ohio State University College of Medicine, Columbus, OH, USA.
[27]Office of Genomic Data Science, National Human Genome Research Institute, Bethesda, MD, USA.
[28]Pacific Northwest Research Institute, Seattle, WA, USA.
[29]Basic Science and Engineering Initiative, Stanford Children's Health, Betty Irene Moore Children's Heart Center, Stanford, CA, USA.
[30]Institute for Stem Cell Biology and Regenerative Medicine, School of Medicine, Stanford University, Stanford, CA, USA.
[31]Division of Genetic Medicine, Department of Pediatrics, University of Washington, Seattle, WA, USA.
[32]Department of Laboratory Medicine and Pathology, University of Washington, Seattle, WA, USA.
[33]Department of Computer Science, Rice University, Houston, TX, USA.
[34]Division of Genetics and Genomics, Boston Children's Hospital, Harvard Medical School, Boston, MA, USA.
[35]Division of Genome Sciences, National Human Genome Research Institute, Bethesda, MD, USA.
[36]Department of Pediatrics, Division of Genetic Medicine, Seattle Children's Hospital, Seattle, WA, USA.
[37]Department of Biomedical Data Science, Stanford University, Stanford, CA, USA.

**Abstract**

Rare diseases are collectively common, affecting approximately one in twenty individuals worldwide. In recent years, rapid progress has been made in rare disease diagnostics due to advances in DNA sequencing, development of new computational and experimental approaches to prioritize genes and genetic variants, and increased global exchange of clinical and genetic data. However, more than half of individuals suspected to have a rare disease lack a genetic diagnosis. The Genomics Research to Elucidate the Genetics of Rare Diseases (GREGoR) Consortium was initiated to study thousands of challenging rare disease cases and families and apply, standardize, and evaluate emerging genomics technologies and analytics to accelerate their adoption in clinical practice. Further, all data generated, currently representing ~7500 individuals from ~3000 families, is rapidly made available to researchers worldwide via the Genomic Data Science Analysis, Visualization, and Informatics Lab-space (AnVIL) to catalyze global efforts to develop approaches for genetic diagnoses in rare diseases (https://gregorconsortium.org/data). The majority of these families have undergone prior clinical genetic testing but remained unsolved, with most being exome-negative. Here, we describe the collaborative research framework, datasets, and discoveries comprising GREGoR that will provide foundational resources and substrates for the future of rare disease genomics.

**Accelerating Diagnoses**

The past decade has seen rapid progress in clinical genetics due to increased discovery of genes and variants involved in Mendelian diseases and ongoing advances in sequencing, variant analysis and data sharing[1–5]. Despite this progress, most individuals who undergo clinical genetic testing for a suspected Mendelian condition remain undiagnosed[6–9]. For example, in the National Human Genome Research Institute (NHGRI) Centers for Mendelian Genomics, while over 3,800 genes were implicated in Mendelian disease, only about 11,000 out of over 28,000 families received a confirmed or potential molecular diagnosis. Thus, significant challenges remain to increase the molecular diagnostic yield and explain currently unsolved rare genetic disorders (**Box 1: Challenges in Diagnosing Rare Genetic Diseases**). In 2021, the NHGRI launched the GREGoR Consortium with five primary research sites and a data coordinating center to accelerate rare disease genetic research by harnessing the latest advances in sequencing including and especially genome sequencing and multi-omics; evaluating and prioritizing the use of functional genomics and novel computational strategies including recent advances in artificial intelligence; translating advances into routine clinical testing; advancing data sharing to foster a quorum of evidence for discovery; and collaborating with rare disease consortium worldwide to continue discovery and reporting of genetic etiologies for Mendelian diseases (**Fig. 1**).

---

**Box 1: Challenges in Diagnosing Rare Genetic Diseases:**

A) The pathogenic variant(s) may be located in a gene yet to be implicated in disease. Until now, over 5,000 protein-coding genes[10,11] have been implicated in at least one disease, but it is estimated that still 10,000+ disease gene relationships[12,13] are undiscovered in just the remaining protein-coding genes.
B) The pathogenic variant(s) may be located in the noncoding genome, where the mechanisms for how a variant manifests a clinical phenotype are not well understood leading to challenges in identifying candidates.
C) The variant may be difficult to detect from solely short-read, exome or genome sequencing such as long repeats, inversions, and complex genomic rearrangements. The variant may be detectable but bioinformatic algorithms may struggle to call the variant correctly such as multi-nucleotide and mosaic variants. The variant may be detectable and called correctly but asserting its functionality or pathogenicity may require unavailable, orthogonal lines of evidence.
D) More complex inheritance patterns such as multi-locus pathogenic variation, oligogenic, polygenic, variable expressivity, incomplete penetrance, imprinting, and/or mosaicism may also be confounding a diagnosis and necessitate a broader approach to understanding the mechanism of disease.
E) A gene-disease relationship may be published or submitted to a genetic database but has yet to be reviewed and incorporated into clinical testing. This is compounded by the rapidly increasing number of Variants of Uncertain Significance (VUS).
F) Many candidate variants and genes are n=1 regardless of the best data sharing practices.
G) Because of the nature of novel discovery, there is not always a functional assay available to provide orthogonal evidence for or against a candidate variant or gene. Many times if a candidate does meet the inclusion criteria for an existing assay, the molecular phenotype measured in the assay may not match the potential mechanism of disease or fully recapitulate the pathophysiological impact, resulting in ambiguous results or an incorrect prediction of pathogenicity.
H) Databases predominantly capture genetic information from individuals of European-like genetic ancestry potentially propagating biases in tools and reference data for variant classification for individuals of non-European-like genetic ancestry[14].
I) Newer genomic technologies may offer advantages over short-read DNA sequencing, but effective and widespread use of these technologies requires clear guidance and broad demonstration of efficacy.

---

**EVALUATING EMERGING METHODS FOR ASCERTAINING RARE DISEASE DIAGNOSES**

**Squeezing the Exome**

The most impactful approach to date for diagnosing rare diseases has been exome sequencing and periodic reanalysis of the protein coding sequences in the human genome[15–19]. GREGoR has led or contributed to 83 papers studying molecular diagnoses in 365 genes with more than a third being novel disease gene discoveries or phenotypic expansions[20–99] (**Supplementary Table 1: Tracking GREGoR Papers With Molecular Diagnoses**) and provided a variety of automated pipelines for large cohort, exome and genome reanalysis, which include phenotypic data integration[16–18]. The success of reanalysis is largely driven by new disease gene discoveries and phenotypic expansions since the original analysis[17], however new tools focused on reanalysis of well-known disease genes and loci have continued to yield diagnostic successes[100]. For example, the inability to phase short-read exome (and also short-read genome) data can confound diagnoses of pathogenic compound heterozygous variants for recessive diseases. To overcome this, GREGoR contributed to a highly accurate method for inferring phase, and has calculated and released all pairwise phasing estimates and usage guidance for rare coding variants in exomes occurring in the same gene through the Genome Aggregation Database (gnomAD)[101,102].

Further, new computational approaches are increasingly able to identify structural variants and copy number variants from exomes. GREGoR researchers have developed tools to identify and implicate hundreds of pathogenic structural variant diagnoses from existing exomes that may have otherwise gone unsolved[38,103–106]. Combined, GREGoR's efforts to continually extract diagnoses from existing exomes demonstrate the ongoing potential for continued innovation in genomic data reanalyses for rare diseases worldwide.

**Short-Read Genome Sequencing**

GREGoR has published a framework guiding usage of genomic technologies when genetic testing via panel or exome sequencing is inconclusive[107]. The next step is typically short-read genome sequencing (srGS). srGS has already become widespread, with large-scale initiatives like the *All of Us* program[108] and UK Biobank[109] releasing srGS for nearly a million individuals. However, similar to exomes, the full diagnostic potential of srGS is still underutilized. For example, the SeqFirst-Neo program[110] using rapid first-line, short-read genome sequencing based on broad eligibility criteria, obtained a precise genomic diagnosis in 50% of infants in an intervention group versus just 10% in the conventional care group. One year later, the intervention group had ninefold greater odds of diagnosis compared to the control group and five times as many infants from underrepresented backgrounds received diagnoses. Ongoing GREGoR research aims to further extract diagnoses from srGS and demonstrate its increasing utility as a first-line clinical test.

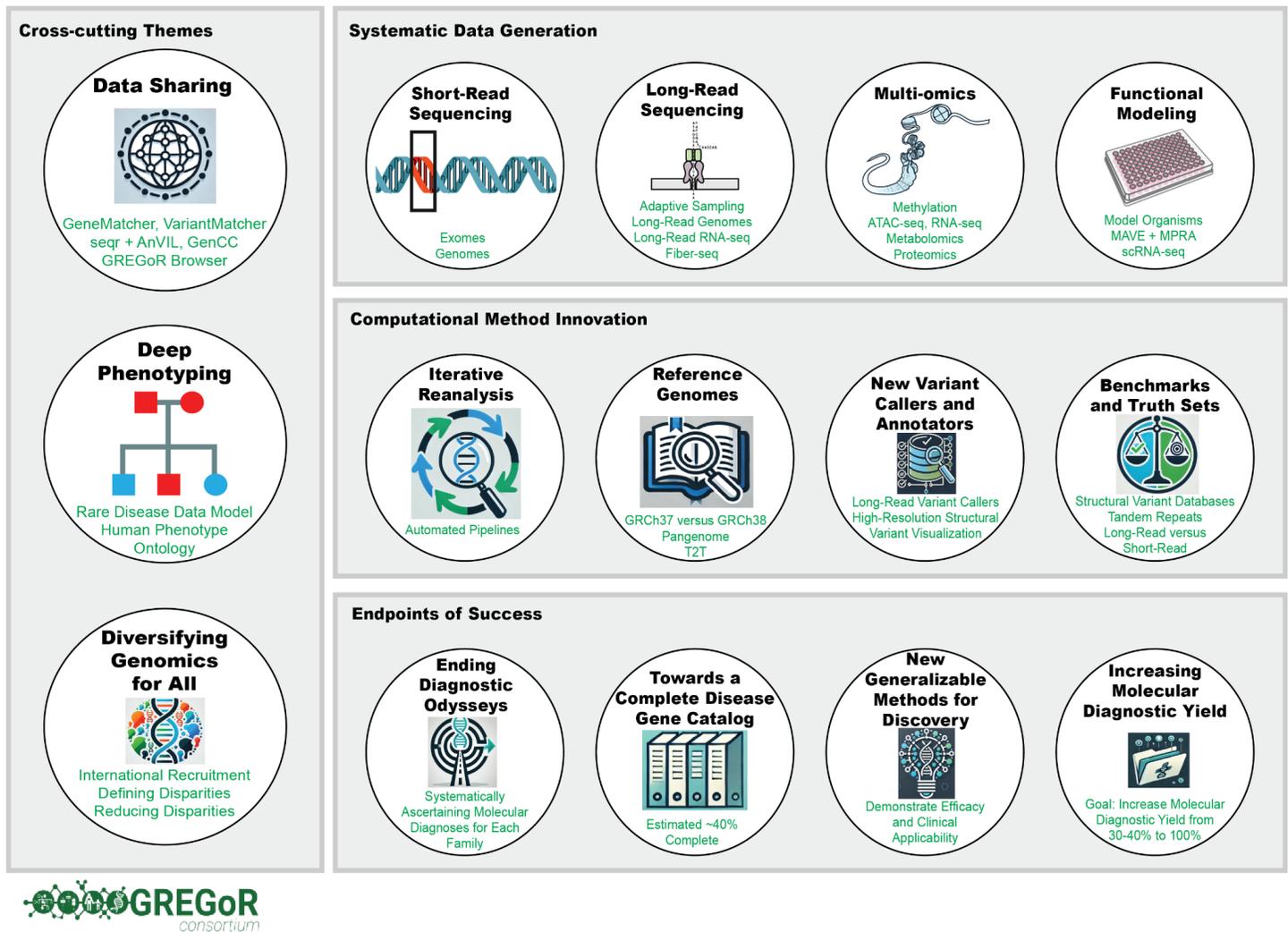

**Figure 1 | Overview of GREGoR.** Strategic framework of the GREGoR Consortium for accelerating genomics in rare disease research, highlighting cross-cutting themes, systematic data generation, computational innovations, and endpoints of success.

However, most molecular diagnoses deduced by srGS are found in protein-coding genes, suggesting they could potentially be detected by exome sequencing. To evaluate the relative utility of srGS, a large-scale study of 822 families by GREGoR researchers reported that of 218 patients who received a diagnosis via srGS, 72% of variants should have been detectable by exome sequencing[111]. The remaining 28% of cases were explained by variants not readily accessible on exomes such as tandem repeat expansions, deep intronic variants, structural variants, and variants in difficult-to-sequence coding regions. Overall, srGS resulted in a >8% increase in the diagnostic yield compared to just exome sequencing and underscored the growing support for using srGS as a first-tier test.

GREGoR is also developing new visualization tools for structural and copy number variants[106] by repurposing read depth data from srGS to mimic SNP arrays to achieve resolution as low as 1 kb - beyond the 5 kb limit of the current standard using array comparative genomic hybridization. Thus, srGS could potentially serve as a cost-effective, first-line, unifying assay by simultaneously replacing both arrays and exomes and enable more accurate, nucleotide-resolution breakpoints of structural variants, which have historically been critical in uncovering mechanisms of genomic rearrangements. Most breakpoints including published structural variants lack validation at nucleotide resolution which is relevant for genomic assembly and hypothesis-driven inferences of SV impact on gene expression. In this same vein, GREGoR investigators are showing that local sequences surrounding candidate and pathogenic variants can offer insights into secondary structure mutagenesis and other mechanisms of genomic disorders[70,71,112]. GREGoR's existing and ongoing work to uncover diagnoses from srGS emphasizes the substantial, yet underutilized, potential of both primary analysis and reanalysis of srGS for rare disease discoveries.

**Long-Read Sequencing**

Long-read sequencing has recently ushered in new diagnostic opportunities in rare diseases. GREGoR and others have demonstrated that use of targeted long-read sequencing can reveal variants, particularly structural variants spanning repetitive sequences, in both known and novel disease genes that are missed or difficult to detect by short-read sequencing[40,97,113–115]. With emerging long-read technologies that allow targeting of specific genomic regions such as adaptive sampling, targeted sequencing panels can evolve beyond coding regions to include UTRs, promoters, intronic, intergenic, and large expanses of noncoding regions around known disease genes. For example, GREGoR in collaboration with Twist developed the Twist Alliance Dark Genes Panel to produce phased variants across 389 medically-relevant and complex autosomal genes, where short-read sequencing tends to fail[116]. Not only were novel pathogenic variants discovered, but an annotated resource was also created to address gaps in current databases for these genes.

One of the major challenges with use of long-read sequencing in rare diseases remains the absence of control datasets for filtering and prioritizing variants. GREGoR is using long-read genome sequencing from individuals with diverse genetic ancestry with an aim to create benchmarks and a database of structural variants for filtering and prioritization, and to catalog structural variants in genes that are difficult to sequence using short-read technology[117–120]. Specifically, GREGoR researchers started the 1000 Genomes Project ONT Sequencing Consortium to generate a database of structural variants derived from long-read sequencing for filtering and prioritization of structural variants in unsolved individuals and recently released the first 100 samples of long-read data from diverse populations[121]. GREGoR is also establishing a baseline catalog of complex structural variants using optical genome mapping. Moreover, GREGoR is developing tools for improved and useful annotation for kilobase and megabase scale variants especially using long-read sequencing with a focus on mosaic structural variants[122,123]. Additionally, innovative computational tools in long-read sequencing variant calling and analysis[123,124] are being developed including *de novo* variant callers and long-read pipelines for mitochondrial variant calling.

As long-read sequencing becomes more commonly used, a key focus for GREGoR is comparing the molecular diagnostic yield and cost-effectiveness of short-read versus long-read sequencing for rare disease cohorts. In multiple studies, long-read genome sequencing uncovered novel candidate variants and genes that were missed by exome or short-read genome sequencing, including *de novo*, compound heterozygous, structural, and epigenetic variants[69,115,125]. In comparison to short-read sequencing, long-read sequencing offers advantages such as better phasing, improved understanding of haplotype blocks, and methylation analysis. GREGoR researchers have been leveraging these advantages by developing tools[126] to utilize methylation data and investigating the diagnostic yield improvements from genome-wide DNA methylation arrays in relation to long-read sequencing[127].

In addition to methylation analyses, usage of long-read sequencing to offer multi-omic insights beyond traditional DNA sequencing are increasingly being applied to find and understand molecular diagnoses as well as mechanisms in rare diseases. One such technology is Fiber-seq, which uses long-read sequencing to simultaneously evaluate primary DNA sequence with a nucleotide-resolution view of surrounding chromatin and epigenetic architecture[128,129]. GREGoR is developing unifying assays looking to simultaneously assay the genome, methylome, epigenome, and transcriptome to identify and understand mechanisms of rare diseases[130]. Such unifying assays help explain previously elusive variants[95] that may have been visible on exome or short-read genome sequencing but may not have been nominated as candidate variants. The comparison and contrast of multiple layers of -omics data in a single, unifying assay mechanistically implicates the pathogenicity of these candidates, especially for noncoding variants.

**Multi-omics**

Complementing advances in DNA sequencing, rapid advances in development of -omics assays continue to provide insights into genome function. In rare disease diagnosis, methylome and transcriptome data have broadly demonstrated their utility by identification of outlier events in methylation[131], splicing, or gene expression implicating pathogenic variants[132–134]. However, there is no standardization of clinical and research transcriptome sequencing despite its promise as a primary diagnostic tool[135]. GREGoR has focused on complementing hundreds of cases with methylation and transcriptome data to facilitate development of standards and new computational methods. For example, recent activities in GREGoR have demonstrated how combined transcriptome and long-read genome analyses can aid in prioritizing structural variants when allele frequency information is limited[136].

An ongoing focus of GREGoR has been creating data that enables evaluation and prioritization of multi-omic assays for rare disease diagnosis. Currently, usage of multi-omics is predominantly limited to research and limited information exists to suggest which post-genome, -omics assay would yield the most useful information. To address this challenge, GREGoR has been generating a squared-off matrix for a subset of families, for whom long-read genome, methylation, chromatin-accessibility, transcriptome, proteome and metabolome data are being collected. Complementing these data, GREGoR has been supporting development and integration of reference -omics data from the Common Fund Data Ecosystem to advance outlier detection for various -omics assays by integrating larger control datasets. Multiple efforts in GREGoR are also facilitating more routine use of these data such as updates to the *seqr* platform to expand intake of multi-omics data types to enable routine linking of outliers in these data to underlying genomic variation[137].

**Functional Modeling**

Functional assays have been critical to identify, screen, and/or validate candidate genes and variants to confirm novel genotype-phenotype relationships and phenotypic expansions and to uncover underlying genomic mechanisms of disease. Of the 83 papers led or facilitated by GREGoR resulting in novel disease gene discoveries or phenotypic expansions, 44 were supported by orthogonal functional experiments to confirm these findings (**[Supplementary Table 1](): Tracking GREGoR Papers With Molecular Diagnoses**). Much of the functional work in the rare disease field has historically relied on classic model organisms like zebrafish, yeast, fruit flies, and mice. Model organism research continues to play a significant role in GREGoR's collaborative efforts with the broader genomics and scientific community, advancing the understanding of disease genes and their variations[138].

An emerging area of technologies for rare disease research is in single cell studies. For example, GREGoR researchers have been able to make high-resolution maps of fetal hematopoiesis to understand how Trisomy 21 predisposes to hematological malignancies[139]; understand regulatory programs contributing to hair and skin diseases[140]; and discover candidate causal variants for Alzheimer and Parkinson disease[141]. Additionally, GREGoR researchers are developing single cell technologies to simultaneously profile chromatin accessibility, transcriptomics, and nuclear protein abundance[142] as well as applying improvements on single-cell, whole genome amplification methods to understand somatic copy number variations in healthy and diseased brain tissue[143].

In addition to single cell and model organism studies, new functional experiments and high-throughput functional genomics are progressively integrated into GREGoR's research through cross-consortium collaborations with the NHGRI IGVF (Impact of Genomic

Variation on Function) consortium[144]. Technologies such as Multiplexed Assays of Variant Effect[145], Massively Parallel Reporter Assays[146], mini-gene splicing assays[147], and high-throughput imaging for cellular mislocalization[148] are providing functional data to validate or invalidate candidate genes and variants and give mechanistic understanding to variant penetrance and pleiotropy[149,150]. For example, GREGoR has exported over 500 candidate genes and variants to IGVF, and more than 100 have been selected for further functional evaluation including in the IGVF's Perturb-seq experimental plans. Further, GREGoR investigators are generating hundreds of isogenic human induced pluripotent stem cell (hiPSC)-derived neural stem cells and glutamatergic induced neurons with CRISPR-engineered, systematic structural variant deletions of local topologically associated domains and chromatin loops[151,152]. These models are eligible to be shared as a GREGoR resource, and in conjunction with transcriptomic and single cell profiles from mice, are being used to study the complex and context-dependent impacts of structural variation on neurodevelopment. Finally, GREGoR is importing predictions of noncoding variant functions from IGVF to improve the understanding of noncoding mechanisms involved in rare diseases[153]. At the same time, GREGoR researchers are advancing the creation and use of developmental cell atlases to develop deep learning models trained on chromatin accessibility and gene expression data representing diverse adult, fetal and developmental contexts to enable the identification of context-specific regulatory effects of rare and *de novo* noncoding variants.

## REFRAMING RARE DISEASE ANALYSIS
### Reference Genomes

Adoption of new reference genomes has lagged in clinical settings. Despite publication of GRCh38 over a decade ago, many clinical labs till today still use GRCh37[154]. Part of the entrenchment of GRCh37 was the lag in necessary infrastructure development to support allele frequencies, *in silico* scores, bioinformatic tools, and clinical databases on GRCh38. Thus, till today, the majority of known clinical disease genes and phenotypic expansions were initially discovered using GRCh37. As a result, research groups, including those in GREGoR, have been studying the differences between GRCh37 and GRCh38 and more references in variant calling and downstream analyses. At the exome level, it has been shown that the reference genome alone impacts variant calling in ~1% of the exome, with 206 genes enriched in discordant calls, including 8 known disease genes[155]. These discrepancies were more pronounced at the RNA level, with research in GREGoR highlighting that 1,492 genes demonstrate reference-dependent quantification, 3,377 genes exhibit reference-exclusive expression, affecting 512 known disease genes[156]. GREGoR investigators have also focused on fixing the GRCh38 reference[157], benchmarking medically relevant genes for both GRCh37 and GRCh38[158], and resolving pathogenic inversions in reference genome gaps using the telomere-to-telomere (T2T) reference[125].

An important question for the field focuses on whether there will be development of flexible pipelines and tools capable of using the newest references, such as T2T and the pangenome[159]. GREGoR in collaboration with Illumina has benchmarked the DRAGEN pipeline which uses graph-based alignment among many other novel features for variant calling in short-read genome sequencing[160]. Looking ahead, GREGoR is collaborating with the Human Pangenome Reference Consortium (HPRC) through methods development like the Pangenome Research Tool Kit[161] to demonstrate accurate variant calling in regions of the genome that were previously too complex for accurate variant calling. Further GREGoR investigators are utilizing pangenome approaches to understand complex tandem repeats in known disease genes[162] and exploring the infrastructure necessary for widespread adoption of the pangenome in clinical settings.

### Deep Phenotyping

Studying phenotypic heterogeneity in the context of genetic heterogeneity is critical to solving unsolved Mendelian disease. Assignment of Human Phenotype Ontology (HPO) terms[163] is a mandatory requirement for GREGoR data collection to allow end users to link all possible genotypes to all possible phenotypes. Additionally, GREGoR is developing novel algorithms based on the directed, acyclic HPO graph to resolve blended phenotypes resulting from multilocus pathogenic variation[8,56], implicate genetic heterogeneity as resulting in similar phenotypic manifestations[49,71], and elucidate gene- and variant-driven phenotypic heterogeneity in rare diseases that have demonstrated genetic heterogeneity[80,164]. Upstream of phenotypic analysis pipelines, GREGoR is building large language models for optimal phenotypic extraction from electronic health records[165]. Understanding phenotypic complexity is particularly critical in tackling the rarest and most challenging molecular diagnoses (**Box 2: The Rarest and Hardest Molecular Diagnoses**), where genetic heterogeneity and diverse mechanisms of pathogenicity require nuanced interpretation and integration of multiple technologies.

---

**Box 2: The Rarest and Hardest Molecular Diagnoses:**

Over time, consistent themes have emerged in the life cycle of disease gene discovery. Typically, the first and easiest candidate variants implicated are *de novo* and/or predicted loss-of-function (pLOF) variants that segregate with the phenotype in a pedigree. These two variant classes have logical frameworks supporting their putative mechanisms. A d*e novo* variant, found in an affected proband but not in the unaffected parents, significantly increases the probability of being causative, especially when other *de novo* cases show the same clinical phenotypes[166,167]. pLOF variants such as nonsense single nucleotide variants, frameshifting insertions or deletions, and splice-site altering variants imply a null effect, because many of these pLOF variants are expected to be caught by the mRNA surveillance mechanism known as nonsense-mediated decay (NMD)[168,169]. NMD is a highly sensitive mechanism present in all tissues and destroys faulty transcripts with premature stops with high efficiency and fidelity[170,171]. This same mechanistic reliability explains why pLOF variants are often the first variant type to be implicated in a novel gene-to-disease discovery, because clinical geneticists can reliably infer that the presence of a pLOF variant will putatively lead to destruction of the faulty RNA transcript and no protein production which is in alignment with a potential pathogenic mechanism of loss of function. Subsequently, other single nucleotide and structural variants in the same gene or region are often implicated after a quorum of cases is established although different types or locations of variation in the same gene can result in distinct clinical phenotypes. Further, once one gene has been implicated it serves as a seed for other genes in the same protein complex[172] or protein pathway[70] or gene family[173,174] to be implicated in the same or similar clinical phenotypes.

Currently, Online Mendelian Inheritance in Man (OMIM) and the Gene Curation Coalition (GenCC) have documented over 5000 genes as being implicated in at least one Mendelian condition[10,11]. Most of these discoveries were achieved through sequencing and interpretation of primary DNA variation in the context of a proband's phenotypes without requiring additional -omics or integrative analyses. While many thousands more disease genes can still be discovered with these same established gene discovery principles, GREGoR is pursuing cases unsolved by standard clinical genetic testing hypothesized to have among the rarest and most difficult-to-detect or interpret pathogenic variation, which many times requires integration of multiple -omics to 1) discover a candidate variant refractory to traditional sequencing methods; 2) provide proper context to interpret a variant that may have been seen in the primary DNA sequence but there was not enough understanding to nominate the variant as a candidate; or 3) provide orthogonal validation of a candidate discovered in the primary DNA sequence but for which the interpretation was speculative at best. Below we discuss 12 of the rarest and hardest molecular diagnoses pursued by GREGoR investigators:

1. **Noncoding variants** occur in genomic regions that do not code for proteins, such as noncoding RNAs, promoters, enhancers, and untranslated regions. Despite not altering protein sequences, these variants can disrupt mechanisms such as gene regulation, splicing, expression, or RNA stability, playing a role in development of rare diseases. Genome sequencing is crucial for detecting variants in noncoding regions, but techniques like ChIP-seq, ATAC-seq, RNA-seq, Fiber-seq and Massively Parallel Reporter Assays can help identify and provide mechanistic explanation for potential pathogenic noncoding variation. For example, noncoding variants such as deep intronic variants may be involved in alterations in splicing, where exons or introns may be incorrectly skipped or included during mRNA processing leading to changes in the final transcript potentially leading to clinical phenotypes[175]. While these variants are often visible in primary DNA sequencing, their interpretation is typically speculative without orthogonal validation such as RT-PCR, RNA sequencing, or mini-gene splicing assays to validate whether or not aberrant splicing occurred by understanding the sequence of the processed mRNA. Similarly, long-read RNA-sequencing is particularly useful for detecting full-length transcripts and interpreting complex splicing patterns. Additionally, splicing-specific variant effect predictors[176] and variant effect predictors that tabulate scores for all the >9 billion single nucleotide variants in both coding and noncoding regions can be deployed to identify potentially pathogenic noncoding variants. Work by GREGoR and others have shown that noncoding variants can be the 'missing variant' in *trans* with a pathogenic coding variant for recessive rare diseases[177,178]. GREGoR is pursuing generalizable methods for understanding noncoding variant effects at scale[44,93].
2. **Noncoding genes** produce molecules that perform potential regulatory, structural, or catalytic roles rather than encode proteins. These include rRNA, tRNA, miRNA, lncRNA, snRNA, and more, and perturbations in noncoding genes may cause rare genetic diseases. A flagship example is perturbations in *RNU4-2* in which cases from multiple GREGoR sites contributed to rapid discovery and progress to publication establishing perturbations in *RNU4-2* as one of the most commonly mutated causes of neurodevelopmental disorders[75,179]. Even though *RNU4-2* is a noncoding RNA, it's discovery timeline is analogous to the typical gene discovery life cycle. The main cases found were *de novo* insertions at the same site in a very large cohort with phenotype-matching cases. *RNU4-2* has now served as a seed for more *RNU4* minor spliceosome genes being implicated in neurodevelopmental disorders[180,181]. Another example from GREGoR is *CHASERR*, a long noncoding RNA adjacent to *CHD2* which was implicated in developmental and epileptic encephalopathy[93]. The *CHASERR* discovery was a strong collaboration with the father of the initial proband serving as a coauthor on the manuscript, highlighting the power of patient partnerships in accelerating rare disease genomics. Further, using Fiber-seq, GREGoR investigators have identified *STRTS*, an intergenic locus implicated in congenital hypothyroidism[95]. These findings illustrate the diagnostic potential of noncoding regions of the genome, which are not systematically included in standard variant analysis workflows. With thousands of noncoding transcripts still poorly understood, GREGoR continues to explore this untapped reservoir of genomic information, paving the way for novel disease-gene discoveries in the noncoding space.
3. **Multilocus Pathogenic Variation (MPV)** refers to the presence of multiple, independently pathogenic variants in multiple genes or loci that collectively contribute to an individual's clinical manifestation[182–184]. These variations can interact in complex ways, leading to compounded effects that may influence disease severity, onset, or progression, often complicating diagnosis and treatment[185]. GREGoR researchers have shown that as many as 5% of individuals where molecular diagnoses are ascertained have MPV[8], with this rate being even higher in rare disease families with parental consanguinity[38,184]. While individual variants comprising MPV are typically routinely diagnosed from just DNA sequencing, the interpretation of MPV typically requires deeper phenotypic analyses and represents a much broader area of gene dosage models for disease causing variation[186].
4. **Oligogenic molecular diagnoses** involve the contribution of variants in two or more genes or loci that collectively lead to a clinical phenotype, a concept distinct from traditional monogenic inheritance patterns or MPV. These cases often present diagnostic challenges because the individual variants may not cause disease independently but act in concert to disrupt pathways or biological networks. GREGoR investigators have shown especially for congenital heart disease the importance of digenic and oligogenic mechanisms[94,182,183]. Currently, we can identify oligogenic variants through DNA sequencing, but the full scope of oligogenic molecular diagnoses may extend across multiple layers of multi-omics. For example, combinations of transcriptomic, proteomic, or metabolomic alterations may converge with DNA variation to create synergistic effects that contribute to disease, representing a new frontier for rare disease genomics.
5. The **Multiple *de novo* Copy Number Variant (M*dn*CNV)** phenotype is a form of multilocus pathogenic variation where four or more independent, constitutional *de novo* copy number variants arise in the same person within one generation[187,188]. GREGoR researchers have shown that this ultra-rare phenotype occurs in roughly 1 in over 12,000 individuals referred for genome-wide chromosomal microarray analysis. M*dn*CNV typically requires integration of multiple technologies as well as use of quantitative phenotyping analysis to fully characterize and understand genomic and clinical impact, including but not limited to arrays, short-read sequencing, and long-read sequencing.
6. **Complex Genomic Rearrangements (CGRs)** are kilobase to megabase-scale structural changes in the genome that involve multiple breakpoints, rearrangements, and/or integration of novel sequences in *cis* that result in duplications, deletions, inversions, and translocations, often affecting gene function and regulation. While CGRs have been catalogued[189–192] and shown to be very abundant across diverse populations, the accurate detection and assembly of CGRs, especially their breakpoints, typically requires more than just short-read DNA sequencing. Long-read sequencing (including adaptive sampling and ultra long-reads), optical genome mapping, chromosomal microarray analysis, and linked-read sequencing can help provide nucleotide resolution for CGRs to understand their mechanisms of formation to better understand how they precipitate genomic disorders and contribute to clinical variability and disease severity [193–195].

7. **Tandem repeats** are sequences where a nucleotide motif is repeated consecutively a varying number of times[196–199]. These repetitive regions can be unstable, leading to expansions or contractions, which are associated with several genetic disorders. Due to the potential for multi-mapped short reads, only shorter tandem repeats have typically been well detected from short-read sequencing technologies. However, long-read sequencing technologies are particularly effective at completely spanning short and longer tandem repeats, enabling accurate determination of repeat length and structure. Additionally, specialized bioinformatic tools[162,200,201] designed for repeat analysis can help in accurately calling tandem repeats and identifying pathogenic expansions or contractions. GREGoR is collating large databases and truth sets of tandem repeats across diverse populations[202–204] to enable more systematic integration of tandem repeat analysis into sequencing pipelines.

8. **Mosaic Variation** originates from post-zygotic mutations and is considered germline if confined to the germ cells or somatic if acquired during or after the first mitotic divisions. Mosaic variation can lead to variations in phenotype, depending on the proportion and distribution of the mutant cells across tissues. Detecting mosaic variation often requires more than standard DNA sequencing due to the low variant allele fraction of mosaic alleles. Reliable detection of mosaic variation typically requires high read-depth sequencing and orthogonal techniques such as digital droplet PCR for validation and bioinformatic discrimination[123] between mosaic variants and sequencing artifacts. GREGoR is collaborating with the SMaHT (Somatic Mosaicism across Human Tissues) consortium to understand pathogenic mosaic variation at scale.

9. **Multi-nucleotide Variants (MNVs)** are two or more variants within the same codon on the same haplotype[205–207]. Accurately identifying and interpreting MNVs is typically a challenge for variant callers and annotation pipelines that may incorrectly interpret the single MNV as multiple independent variants. However, it has been shown that there are over 50 variants per person affected by MNVs[208]. Incorrect interpretation of MNVs can alter the interpretation of clinically pathogenic variation such as nonsense single nucleotide variants that do not lead to a premature stop codon introduction. Many variant effect predictors and Multiplex Assays of Variant Effects systematically score every possible MNV in a target locus. GREGoR is evaluating the utility of these data towards VUS reclassification and novel disease gene discovery.

10. **NMD Escaping Variants** introduce premature stop codons in mRNA that evade NMD, producing truncated proteins of unknown function that may or may not manifest clinical disease. These variants are visible on primary DNA sequencing and because extensive work has been done to determine the rules of NMD escape[209,210], many of these variants are already implicated in clinical disease and can be speculated on from just primary DNA sequencing[171,211]. However, newer approaches such as long read RNA-seq and proteomics are opening new doors into the investigation of NMD escape alleles and their downstream mechanisms.

11. **Incompletely Penetrant Variants** refer to pathogenic changes in DNA that do not always result in observable clinical disease, even in individuals carrying the variant. This variability can complicate interpretation, especially in family studies where carriers appear unaffected[68,71,212]. These variants often challenge diagnostic workflows because traditional penetrance assumptions do not hold, necessitating integration of orthogonal lines of evidence such as animal models[71,212] and epigenetics[89] to understand the variant pathogenicity. These functional efforts are particularly critical for diseases where penetrance may be age-dependent, sex-influenced, or modified by external factors[89]. Further, in a case-by-case analysis across all of gnomAD v4, >95% of incompletely penetrant pLOF variants found in severe, early-onset, highly penetrant haploinsufficient disease had explainable causes such as a downstream frame-restoring variant, predicted re-initiation by a downstream methionine, a MNV changing the interpretation of a nonsense variant to a missense or synonymous variant, or the location of the pLOF variant being in an NMD Escape region[213].

12. **Variants of Uncertain Significance (VUS)** are most often reported from testing genes that already have been established in disease pathogenesis[214], though by definition, all variants found in candidate genes with insufficient evidence for disease implication are also VUS. VUS are accumulating rapidly over time as testing volume expands. In fact, VUS are more often reported during panel testing compared to exome or genome sequencing due to professional practices[215] and are disproportionately called in individuals from non-European-like genetic ancestry[14]. Thus, while transitioning to consistent first-line clinical usage of exome or genome sequencing coupled with phenotypic analyses will decrease the rate of clinically reported VUS, given that most causal variants are only found in a single individual, integration of functional modeling is often required to reclassify VUS. Multiplexed Assays of Variant Effects are high-throughput experiments orthogonal to the clinical sequencing pipeline that produce functional scores for all variant effects in a target locus and when their evidence strength is clinically calibrated and incorporated with other lines of evidence, demonstrate significant promise in massive VUS reclassification[14,216].

## DIVERSIFYING RARE DISEASE GENOMICS AND DIAGNOSTICS

Participation in rare disease research can be influenced by numerous factors, including but not limited to institutional, socioeconomic, geographic, linguistic, cultural, educational, and insurance factors[217–220]. GREGoR sites, many of which are in urban centers, have implemented procedures for online enrollment, chatbots[221], remote consent, offsite sample collection (including mobile phlebotomy in rural areas), and translated materials in multiple languages to improve accessibility. Attention to these details has allowed GREGoR to foster many international collaborations with local scientists to sequence and make available sequencing data from thousands of individuals of non-European-like genetic ancestry, actively seeking to address the disparities in genomic data availability across Middle Eastern, North African, Southeast Asian, South American, and other underrepresented groups such as African-American and Hispanic peoples.

Diversifying participants in genomics research via recruitment of participants from underrepresented populations is just one approach to fostering equity. GREGoR is pursuing orthogonal approaches to increase access to a genetic diagnosis by pursuing improved variant calling methods, applying multiplexed functional assays to improve interpretation of VUS that are enriched in underrepresented populations, and testing technologies and workflows that reduce barriers to equitable access to a genetic diagnosis. For example, GREGoR is collaborating with the Human Pangenome Reference Consortium to reduce bias and improve variant calling accuracy across all populations via usage of the pangenome. GREGoR has analyzed population biobank data to show a higher prevalence of Variants of Uncertain Significance and fewer Pathogenic or Likely Pathogenic classifications in individuals of non-European-like genetic ancestry[14]. These disparities were alleviated by using high throughput, multiplexed functional experiments to test every possible single variant in genes of interest to resolve VUS disparities between populations. However, this study demonstrated that allele frequency and variant effect predictors contribute to the inequitable classification of variants and more work to prevent bias in clinical variant classification is an important future priority for GREGoR. Finally, the SeqFirst program[110] shows that using simple criteria to assess eligibility for rapid short-read genome sequencing significantly increases the proportion of non-White and Black infants who receive a precise genetic diagnosis. GREGoR and SeqFirst are conducting a comparison of long-read versus short-read sequencing in the SeqFirst cohort to better understand the relative value of these technologies within populations in this cohort.

**ACCELERATING DATA SHARING**

Data sharing is critical to the advancement of rare disease genomics and diagnoses[222,223]. GREGoR is committed to rapid release of genomic and phenotype data to the larger research community within AnVIL and making these data FAIR (Findable, Accessible, Interoperable, and Reusable)[224] and machine-readable. At the time of AnVIL submission and prior to analysis, deep phenotyping (**Fig. 2A**), potentially multiple orthogonal lines of -omics data (**Fig. 2B, 2C**), and pedigree data (**Fig. 2D**) are made available via the GREGoR data model for broader dissemination to the research community. Currently, DNA data on approximately 7400 individuals from over 3000 families is available with transcriptome data available for over 500 individuals and nearly 200 participants with both exome and short-read genome data (dbGaP:phs003047; **Fig. 2B**). Within the next year, planned releases include additional short and long read genomes, short and long read RNA-seq, Fiber-seq, ATAC-seq, metabolomics and proteomics. GREGoR has begun to assess cases (**Fig. 2E**) and has identified candidate discoveries and molecular diagnoses for over 400 families which are added in subsequent AnVIL submissions over time. GREGoR hopes that solved cases can be used by the community as positive controls for benchmarking tools and that the current dataset of unsolved cases (many of which are exome negative) holds tremendous potential for discovery by collaborative community efforts.

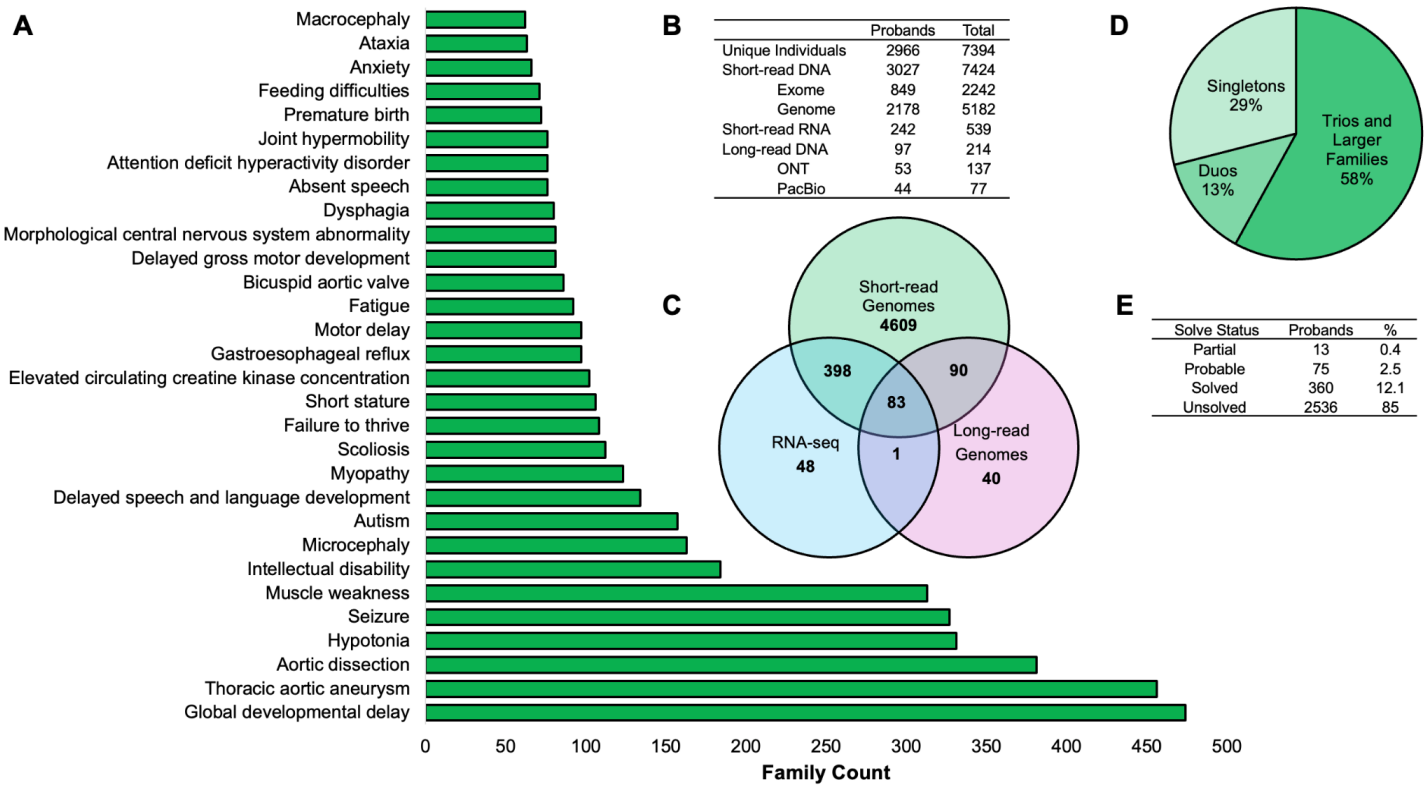

**Figure 2 | Overview of Publicly Released GREGoR Data.** Summary of second public data release (dbGaP:phs003047). (A) Distribution of top 30 phenotypes in GREGoR based on Human Phenotype Ontology descriptions. (B) Table of numbers for probands and total individuals for each sequencing modality. (C) Venn diagram depicting overlap across short-read genomes, RNA-seq, and long-read genomes in data generation. (D) Family structures comprising the overall cohort from a total of n=3059 families. (E) Summary of current solved cases. Data is shared prior to analysis but even the current diagnostic outcomes underscore the challenges and opportunities in resolving rare disease cases that are previously exome negative.

Many commercial and academic clinical labs have undiagnosed cases potentially explained by novel gene discoveries or phenotypic expansions. The GREGoR Consortium has actively engaged with clinical labs to study effective strategies for exome and genome analysis without overburdening variant analysts[225]. GREGoR has released recommendations for clinical labs to report variants in novel candidate genes and support follow-up investigations, enabling broad discoveries and patient diagnoses[226].

To enable exchange of data and facilitate collaboration, all GREGoR candidate genes are shared to Matchmaker Exchange[227,228] through either GeneMatcher[212], *seqr*[130] or MyGene2[229]. Notably, for the 83 GREGoR publications (**[Supplementary Table 1](#): Tracking GREGoR Papers With Molecular Diagnoses**) involving novel disease genes or phenotypic expansions, almost every project has been influenced by findings from connections made across or within nodes of the Matchmaker Exchange. Novel candidate genes and phenotypic expansions are also curated for validity and publicly shared to the Gene Curation Coalition database to accelerate access to early evidence of novel gene-disease relationships and aid in standardized clinical diagnostics and research[11]. Analogously, candidate variants and molecular diagnoses are deposited in ClinVar[230,231]. Also GREGoR is leveraging federated variant-level matchmaking through tools such as VariantMatcher[232,233], which allows queries of variant data across different genomic datasets, even when a variant or gene has not been recognized as a candidate with the hope of accelerating disease-causing variant-level discovery within and beyond the exome.

GREGoR has developed a novel data model that emphasizes the essentials of rare disease research such as the importance of data accessibility, consent consistency and transparency, and the usage of many accepted ontologies and common standards. Every variant in the GREGoR joint callset is machine-readable with both a unique GA4GH Variant Representation Specification (VRS) ID[234] and ClinGen allele ID[235], and the data model accommodates variants and output files from a wide variety of genomic, multi-omic, phenotypic, and molecular data types, and is modularly designed to support the integration of future data types. GREGoR implements this model within AnVIL workspaces for data submission and validation to enable (1) streamlined genomic and phenotypic data deposition; (2) scalable, semi-automated quality control of molecular, phenotypic, and variant annotation data; and (3) expedited, controlled-access release to the broader research community.

Inside AnVIL, the GREGoR data is also queryable via *seqr*, which integrates variant filtration, annotation, and causal variant identification[137]. Outside AnVIL, GREGoR has developed a public variant browser[236], which already includes >95 million variants. Not only is the number of families predicted to triple by the end of GREGoR, but de-identified phenotypic data are also being added to the public browser, allowing researchers to easily explore putative genotype-phenotype relationships in rare disease families.

**CONCLUSION**

GREGoR aims to advance state-of-the-art approaches to determine molecular diagnoses for individuals with unsolved rare diseases. Central to this approach is the creation of a broadly accessible and information-rich genomics data resource derived from individuals and families with rare diseases for whom prior standard-of-care testing such as exome sequencing has been non-diagnostic. This resource is defined by a future-forward data model and infrastructure that incorporates genomic and other -omic datasets generated through emerging technologies, family structure data, and rich phenotypic data and currently is supporting data for over 3500 families, many of whom remain unsolved.

Several opportunities remain to advance rare disease research and accelerate diagnoses. Foremost, the effort to complete a comprehensive catalog of genes underlying Mendelian conditions remains far from finished. While thousands of Mendelian conditions have been described, a significant proportion of these conditions still lack a known genetic cause, leaving substantial gaps in our understanding. Further, even when all genes for Mendelian conditions are identified, causal variant discovery will remain far from saturated. This is particularly true for missense variants, which often require functional validation, and for noncoding variants, where the regulatory mechanisms are complex and poorly characterized. Far from being a task of "wrapping up the edges," these challenges represent a vast forefront in genomic research, demanding both innovative methodologies and sustained collaboration to make meaningful progress. Alongside these challenges, the advancement of functional genomic assays has required the vetting of these approaches at scale in individuals of diverse genetic ancestries and diverse rare disease phenotypes. Such efforts are critical to establishing the standards of when and how to use a specific approach and will guide expectations on their relative yields at scale and their adoption in clinical practice. Lastly, there is a palpable need to translate scientific discoveries into curation practices that align with formal clinical standards. To address these gaps, GREGoR provides data and infrastructure that will catalyze the development and implementation of new approaches to advance genomics in rare disease by the broader community.

**Acknowledgements:**
We would like to acknowledge all patient participants and their families. We would also like to acknowledge the expansive set of collaborators including clinical providers, analysts and rare disease researchers. This work was supported by the NIH NHGRI GREGoR Consortium (U01HG011758, U01HG011755, U01HG011762, U01HG011745, U01HG011744, U24HG011746).

**Author contributions:**
Conceptualization, Methodology: MD, BH, MMW, RAU, JLT, LW, SB, JAB, JXC, ECD, EEE, RAG, JRL, AS, MET, AHW, CLW, CW, MTW, CMBC, CAG, SM, DEM, HLR, FJS, EV, AODL, JEP, LHC, MJB, SBM; Data curation: MD, BH, MMW, SBM; Investigation: SB, JAB, JXC, ECD, EEE, RAG, JRL, AS, MET, AHW, CLW, CW, MTW, CMBC, CAG, SM, DEM, HLR, FJS, EV, AODL, JEP, LHC, MJB, SBM; Data curation: MD, BH, MMW, SBM; Project administration: SBM; Writing - original draft: MD, BH, MMW, RAU, JLT, LW, CMBC, CAG, SM, DEM, HLR, FJS, EV, AODL, JEP, LHC, MJB, SBM; All authors contributed to developing and enacting the vision and goals of the GREGoR Consortium, reviewing, and editing the manuscript.

**Competing Interests:**
All authors of this manuscript are funded by the NIH and NHGRI. JRL has stock ownership in 23andMe, is a paid consultant for Regeneron Genetics Center, and is a co-inventor on multiple U.S. and European patents related to molecular diagnostics for inherited neuropathies, eye diseases, and bacterial genomic fingerprinting. JRL serves on the Scientific Advisory Board of Baylor Genetics. JRL and RAG declare that Baylor Genetics is a Baylor College of Medicine affiliate that derives revenue from genetic testing. BCM and Miraca Holdings have formed a joint venture with shared ownership and governance of Baylor Genetics which performs clinical microarray analysis and other genomic studies (exome and genome sequencing) for patient and family care. FJS has received research support from Illumina, Pacific Biosciences, and Genentech. JEP is an advisor to MaddieBio. SBM is an advisor to BioMarin, MyOme, and Tenaya Therapeutics. FJS, DEM have received research support and/or consumables from ONT and have received travel funding to speak on behalf of ONT. DEM has received travel support from Pacific Biosciences. DEM is on an advisory board at ONT, a scientific advisory board at Basis Genetics, and holds stock options in both MyOme and Basis Genetics. MJB is the chair of the Scientific Advisory Board of GeneDx and receives funding from the American Society of Human Genetics as the Editor-in-Chief of HGG Advances. JXC receives funding from the American Society of Human Genetics as the Deputy Editor of HGG Advances. DP consults for Ionis Pharmaceuticals. HLR has received rare-disease research funding from Microsoft and Illumina and compensation as a past member of the scientific advisory board of Genome Medical. AODL was a paid consultant to Tome Biosciences, Ono Pharma USA, Addition Therapeutics, Congenica and receives research funding from Pacific Biosciences. EEE is a scientific advisory board member of Variant Bio, Inc. and is an investigator of the Howard Hughes Medical Institute.

**GREGoR Banner Authors**:
**U01HG011758**
Jennifer Posey, Richard Gibbs, James Lupski, Hatoon Al Ali, Elizabeth Atkinson, Sairam Behera, Shaghayegh Beheshti, Eric Boerwinkle, Tugce Bozkurt-Yozgatli, Daniel Calame, Ivan Chinn, Zeynep Coban-Akdemir, Karen Coveler, Zain Dardas, Moez Dawood, Harsha Doddapaneni, Haowei Du, Ruizhi Duan, Iman Egab, Jawid Fatih, Mira Gandhi, Brandon Garcia, Nikhita Gogate, Christopher Grochowski, Jianhong Hu, Minal Jamsandekar, Shalini Jhangiani, Angad Jolly, Parneet Kaur, Ahmed K. Saad, Jesse Levine, Richard Lewis, Yidan Li, Pengfei Liu, Medhat Mahmoud, Dana Marafi, Tadahiro Mitani, Chloe Munderloh, Donna Muzny, Sebastian Ochoa Gonzalez, Piyush Panchal, Shruti Pande, Davut Pehlivan, Archana Rai, Edgar Andres Rivera-Munoz, Aniko Sabo, Evette Scott, Fritz Sedlazeck, V. Reid Sutton, Kim Walker, Lauren Westerfield, Jiaoyang Xu, Bo Yuan, Xinchang Zheng

**U01HG011755**
Anne O'Donnell-Luria, Heidi Rehm, Michael Talkowski, Siwaar Abouhala, Kaileigh Ahlquist, Mutaz Amin, Christina Austin-Tse, Samantha Baxter, Benjamin Blankenmeister, Philip Boone, Harrison Brand, Colleen Carlston, Celine de Esch, Stephanie DiTroia, Michael Duyzend, Vijay Ganesh, Kiran Garimella, Carmen Glaze, Emily Groopman, Sanna Gudmundsson, Stacey Hall, Yongqing Huang, Julia Klugherz, Katie Larsson, Arthur Lee, Gabrielle Lemire, Jialan Ma, Daniel MacArthur, Brian Mangilog, Daniel Marten, Eva Martinez, Olfa Messaoud, Mariana Moyses, Ashana Neale, Emily O'Heir, Melanie O'Leary, Ikeoluwa Osei-Owusu, Lynn Pais, Alicia Pham, Lindsay Romo, Kathryn Russell, Monica Salani, Kaitlin Samocha, Alba Sanchis-Juan, Jillian Serrano, Gulalai Shah, Moriel Singer-Berk, Mugdha Singh, Hana Snow, Kayla Socarras, Sarah Stenton, Jui-Cheng Tai, Grace VanNoy, Ben Weisburd, Michael Wilson, Monica Wojcik, Isaac Wong, Rachita Yadav

**U01HG011762**
Stephen Montgomery, Jon Bernstein, Matthew Wheeler, Emily Alsentzer, Raquel Alvarez, Euan Ashley, Themistocles Assimes, Gill Bejerano, Devon Bonner, Denver Bradley, Jennefer Carter, Clarisa Chavez, Ziwei Chen, Salil Deshpande, Sara Emami, Ivy Evergreen, Casey Gifford, Pagé Goddard, John Gorzynski, William Greenleaf, Rodrigo Guarischi-Sousa, Caitlin Harrington, Sohaib Hassan, Tanner Jensen, David Jimenez-Morales, Christopher Jin, Aimee Juan, Jessica Kain, Laura Keehan, Anshul Kundaje, Soumya Kundu, Samuel M. Lancaster, Shruti Marwaha, Dena Matalon, Taylor Maurer, Lauren Meador, Hector Rodrigo Mendez, Alexander Miller, Matthew B. Neu, Thuy-mi P. Nguyen, Jonathan Nguyen, Jeren Olsen, Evin Padhi, Paul Petrowski, Astaria Podesta, Elizabeth Porter, Wanqiong Qiao, Thomas Quertermous, Chloe Reuter, Oriane Rubio, Stuart Scott, Riya Sinha, Kevin S. Smith, Michael Snyder, Brigitte Stark, Suchitra Sudarshan, Christina Tise, Philip Tsao, Rachel Ungar, Isabella Voutos, Juliana Walrod, Ziming Weng, Laurens Wiel, Frank Wong, Yao Yang, Jiye Yu, Jimmy Zhen

**U01HG011745**
Eric Vilain, Seth Berger, Emmanuèle Délot, Miguel Almalvez, Rishi Aryal, Light Auriga, Rebekah Barrick, Sami Belhadj, Krista Bluske, Leandros Boukas, Andrea J. Cohen, Ya Cui, Ivan de Dios, Meghan Delaney, Jamie Fraser, Vincent Fusaro, John Harting, Megan Hawley, Yun-Hua Hsiao, Amanda Kahn-Kirby, Rachid Karam, Charles Hadley King, Arthur Ko, Wei Li, Bojan Losic, Jonathan LoTempio, Sofia Marmolejos, Robert Nussbaum, Georgia Pitsava, Sarah Savage, Emily Westheimer, Changrui Xiao, Jianhua Zhao

**U01HG011744**
Michael Bamshad, Chia-Lin Wei, Evan Eichler, Jessica Chong, Kailyn Anderson, Peter Anderson, Sabrina Best, Elizabeth Blue, Kati Buckingham, Silvia Casadei, Yong-Han Cheng, Colleen Davis, Sophia Gibson, William Gordon, Jonas Gustafson, William Harvey, Martha Horike-Pyne, Gail Jarvik, Annelise Mah-Som, Colby Marvin, Francesco Kumara Mastrorosa, Sean McGee, Heather Mefford, Danny Miller, Miranda Zalusky, Karynne Patterson, Matthew Richardson, Adriana Estela Sedeño-Cortés, Joshua Smith, Olivia Sommerland, Lea Starita, Andrew Stergachis, Elliott Swanson, Jeffrey Weiss, Qian Yi, Christina Zakarian

**U24HG011746**
Susanne May, Ali Shojaie, Emily Bonkowski, Sarah Conner, Matthew Conomos, Stephanie Gogarten, Ben Heavner, Sarah Nelson, Sheryl Payne, Jaime Prosser, Guanghao Qi, Adrienne Stilp, Catherine Tong, Marsha Wheeler, Quenna Wong

**GREGoR Partner Members**
Aashish Adhikari, Kinga Bujakowska, Claudia M. B. Carvalho, Ali Crawford, Aimée M. Dudley, Kelly Hagman, Yang I. Li, Jill Moore, Aaron R. Quinlan, Alex Wagner, Bo Xia, S. Stephen Yi

**NHGRI**
Lisa Chadwick, Christopher Wellington, Sara Currin, Gaby Villard

**Writing Group**
Moez Dawood, Ben Heavner, Marsha Wheeler, Rachel A. Ungar, Jonathan LoTempio, Laurens Wiel, Claudia M. B. Carvalho, Casey A. Gifford, Susanne May, Danny E. Miller, Heidi L. Rehm, Fritz J. Sedlazeck, Eric Vilain, Anne O'Donnell-Luria, Jennifer E. Posey, Lisa H. Chadwick, Michael J. Bamshad, Stephen B. Montgomery

**Supplementary Table 1: Tracking GREGoR Papers With Molecular Diagnoses**

| Gene | PMID | Functional Work |
|---|---|---|
| ACBD6 | 34582790 | No |
| ACBD6 | 37951597 | Yes |
| ACCN2 | 34582790 | No |
| ACOT7 | 34582790 | No |
| ACTC1 | 37457373 | No |
| ACTL6A | 34582790 | No |
| ACTL6B | 39275948 | Yes |
| ACTR1B | 39033378 | No |
| ADAM19 | 34582790 | No |
| ADAMTS15 | 35962790 | Yes |
| ADSL | 34582790 | No |
| AFF3 | 38811945 | Yes |
| AHDC1 | 34582790 | No |
| AHDC1 | 33372375 | No |
| AHDC1 | 34950897 | No |
| ALS2 | 34582790 | No |
| AMPD2 | 34582790 | No |
| ANK3 | 38988293 | No |
| ANKRD11 | 34582790 | No |
| AP3B2 | 34582790 | No |
| AP4B1 | 34582790 | No |
| APTX | 34582790 | No |
| ARAP1 | 34582790 | No |
| ARFGEF3 | 38258669 | Yes |
| ARID1B | 34582790 | No |
| ARID4A | 34582790 | No |
| ARV1 | 34582790 | No |
| ARX | 34582790 | No |
| ASH1L | 34582790 | No |
| ASNS | 34582790 | No |
| ASPM | 34582790 | No |
| ASTN1 | 34582790 | No |
| ASTN2 | 34582790 | No |
| ASXL3 | 34582790 | No |
| ATP1A1 | 34582790 | No |
| ATP1A3 | 37043503 | Yes |
| ATP5F1A | 34954817 | Yes |
| ATP5F1E | 34954817 | Yes |
| ATP5MC3 | 34954817 | Yes |
| ATP5PO | 34954817 | Yes |
| ATP7A | 34582790 | No |
| ATRX | 34582790 | No |
| BARD1 | 34582790 | No |
| BHLHA9 | 36035248 | No |
| BMPER | 34582790 | No |
| BRWD3 | 34582790 | No |

| Gene | PMID | Novel |
|---|---|---|
| C2ORF69 | 34582790 | No |
| CACNA2D2 | 34582790 | No |
| CAMSAP1 | 34582790 | No |
| CAPN3 | 34816580 | No |
| CASP5 | 37603195 | Yes |
| CBX6 | 34582790 | No |
| CC2D1B | 34582790 | No |
| CCDC39 | 39606420 | No |
| CCDC40 | 39606420 | No |
| CCNO | 39606420 | No |
| CDK10 | 34582790 | No |
| CDKL5 | 35934918 | No |
| CDKL5 | 34582790 | No |
| CELF2 | 38258669 | Yes |
| CELSR3 | 38429302 | Yes |
| CEP290 | 34582790 | No |
| CEP85L | 34582790 | No |
| CFAP46 | 39606420 | No |
| CFAP47 | 38633811 | Yes |
| CHASERR | 39442041 | Yes |
| CHD2 | 39442041 | Yes |
| CHD3 | 34582790 | No |
| CHMP1A | 34582790 | No |
| CIT | 34582790 | No |
| CLP1 | 34582790 | No |
| CNTN5 | 34582790 | No |
| CNTNAP2 | 34582790 | No |
| COBL | 34582790 | No |
| COG3 | 37711075 | Yes |
| COL6A1 | 38585825 | Yes |
| COPB1 | 34582790 | No |
| CREB3 | 34582790 | No |
| CRYBB2 | 38990107 | No |
| CTBP1 | 34582790 | No |
| CTNNA1 | 38585811 | No |
| CUX1 | 38585811 | No |
| CYP1B1 | 38990107 | No |
| DCX | 34582790 | No |
| DDC | 34582790 | No |
| DDX3X | 34582790 | No |
| DEAF1 | 34582790 | No |
| DENND5A | 39174524 | Yes |
| DHCR24 | 34582790 | No |
| DHX9 | 37467750 | Yes |
| DLGAP1 | 34582790 | No |
| DNAAF2 | 39606420 | No |
| DNAAF4 | 39606420 | No |
| DNAH1 | 39606420 | No |

| Gene | PMID | Novel |
|------|------|-------|
| *DNAH1* | 39606420 | No |
| *DNAH11* | 39606420 | No |
| *DNAH5* | 39606420 | No |
| *DNAH6* | 39606420 | No |
| *DNAH6* | 39606420 | No |
| *DNAH8* | 39606420 | No |
| *DNAH9* | 39606420 | No |
| *DNAJC8* | 34582790 | No |
| *DPF2* | 34582790 | No |
| *DRC1* | 39606420 | No |
| *DUSP4* | 34582790 | No |
| *DYRK1A* | 34582790 | No |
| *EEF1A2* | 34582790 | No |
| *ELF4* | 36477361 | Yes |
| *ENPP6* | 34582790 | No |
| *ENTPD1* | 35471564 | Yes |
| *EPG5* | 34582790 | No |
| *EPHA8* | 34582790 | No |
| *ERCC6* | 34582790 | No |
| *ESAM* | 34582790 | No |
| *ESAM* | 36996813 | Yes |
| *EXOSC3* | 34582790 | No |
| *FA2H* | 34582790 | No |
| *FAM120A* | 34582790 | No |
| *FAM91A1* | 34582790 | No |
| *FBP2* | 38258669 | Yes |
| *FBXW11* | 34582790 | No |
| *FER* | 38585811 | No |
| *FGF21* | 38585811 | No |
| *FLNA* | 39606420 | No |
| *FLVCR1* | 39306721 | Yes |
| *FOXG1* | 34582790 | No |
| *FOXI3* | 37041148 | Yes |
| *FOXN4* | 34582790 | No |
| *FRMD7* | 34582790 | No |
| *FSHR* | 34582790 | No |
| *GAS8* | 39606420 | No |
| *GATSL3* | 34582790 | No |
| *GCC2* | 34582790 | No |
| *GCH1* | 35083481 | No |
| *GGPS1* | 35869884 | No |
| *GIN1* | 34582790 | No |
| *GIPR* | 34582790 | No |
| *GIT1* | 34582790 | No |
| *GJC2* | 34582790 | No |
| *GLB1* | 34582790 | No |
| *GLI2* | 38990107 | No |
| *GLI2* | 34582790 | No |

| Gene | PMID | Result |
|---|---|---|
| *GLI3* | 36035248 | No |
| *GMPPB* | 34582790 | No |
| *GNAS* | 35811283 | No |
| *GNAS* | 38585811 | No |
| *GOLGA2* | 34582790 | No |
| *GOLGA4* | 34582790 | No |
| *GPR87* | 36622818 | No |
| *GPT2* | 34582790 | No |
| *GREB1L* | 37124138 | No |
| *GRM7* | 34582790 | No |
| *HECTD3* | 34582790 | No |
| *HECTD4* | 34582790 | No |
| *HECTD4* | 36401616 | No |
| *HEXB* | 34582790 | No |
| *HMGCR* | 37167966 | Yes |
| *HOXD* | 36035248 | No |
| *HOXD13* | 36035248 | No |
| *HPDL* | 34582790 | No |
| *HPS1* | 34582790 | No |
| *HSPB1* | 33686258 | No |
| *HYAL2* | 34906488 | Yes |
| *HYDIN* | 39606420 | No |
| *ITGB8* | 34582790 | No |
| *JRK* | 34582790 | No |
| *KANSL3* | 38258669 | Yes |
| *KCNJ14* | 34582790 | No |
| *KCTD7* | 34582790 | No |
| *KDM2B* | 34582790 | No |
| *KDM5A* | 34582790 | No |
| *KDM5B* | 39202393 | No |
| *KDM5C* | 34582790 | No |
| *KIAA0430* | 34582790 | No |
| *KIF1A* | 34582790 | No |
| *KIF21A* | 38585811 | No |
| *KIF26A* | 34582790 | No |
| *KIF26A* | 36228617 | Yes |
| *KIF5C* | 38585811 | No |
| *KIF7* | 39606420 | No |
| *KIFC3* | 34582790 | No |
| *KLB* | 38585811 | No |
| *L1CAM* | 34582790 | No |
| *LAMA1* | 34582790 | No |
| *LAMB3* | 34582790 | No |
| *LAMC3* | 34582790 | No |
| *LARGE1* | 34582790 | No |
| *LARP7* | 34582790 | No |
| *LCTL* | 34582790 | No |
| *LGI3* | 35948005 | Yes |

| Gene | PMID | Flag |
|---|---|---|
| LPAR6 | 34582790 | No |
| LRP2 | 34582790 | No |
| LSS | 37157980 | No |
| MAP2K4 | 38258669 | Yes |
| MAP3K20 | 38451290 | No |
| MAP3K7 | 34582790 | No |
| MCM3AP | 34582790 | No |
| MCM6 | 38258669 | Yes |
| MCPH1 | 34582790 | No |
| MDM1 | 38868186 | Yes |
| MEGF8 | 39606420 | No |
| MGAT2 | 34582790 | No |
| MGP | 37923733 | Yes |
| MKS1 | 34582790 | No |
| MPZ | 38585811 | No |
| MRPS25 | 34582790 | No |
| MRPS25 | 39606420 | No |
| MRPS27 | 34582790 | No |
| MTOR | 34582790 | No |
| MTSS2 | 36067766 | Yes |
| MUSK | 34816580 | No |
| MYH1 | 34582790 | No |
| MYH10 | 38585811 | No |
| NALCN | 34582790 | No |
| NANS | 34582790 | No |
| NAV2 | 34582790 | No |
| NES | 38585811 | No |
| NETO1 | 36622818 | No |
| NFE2L3 | 38258669 | Yes |
| NGEF | 34582790 | No |
| NGLY1 | 34582790 | No |
| NHLRC2 | 37188825 | Yes |
| NLK | 34582790 | No |
| NODAL | 38570875 | No |
| NOTCH1 | 33686258 | No |
| NPHP3 | 39606420 | No |
| NPR2 | 36035248 | No |
| NRD1 | 34582790 | No |
| NSD1 | 34582790 | No |
| NTNG2 | 34582790 | No |
| NUAK1 | 34582790 | No |
| OCLN | 34582790 | No |
| OLIG2 | 38585811 | No |
| OTOA | 33492714 | No |
| OTUD6B | 35430327 | No |
| PAFAH1B1 | 34582790 | No |
| PARD3B | 34582790 | No |
| PAX5 | 35094443 | No |

| Gene | PMID | Included |
|---|---|---|
| PCDH18 | 34582790 | No |
| PDK1L1 | 39606420 | No |
| PDZD2 | 34582790 | No |
| PEX6 | 34582790 | No |
| PGAP3 | 34582790 | No |
| PHF8 | 35469323 | No |
| PIK3C2A | 34582790 | No |
| PKD1 | 34582790 | No |
| PLAA | 34582790 | No |
| PLCG1 | 38260438 | Yes |
| PLD3 | 34582790 | No |
| PLK4 | 34582790 | No |
| PLXNA1 | 34582790 | No |
| PNKP | 34582790 | No |
| POLR1D | 34582790 | No |
| POLR3A | 34582790 | No |
| POMGNT1 | 34582790 | No |
| PPP1R15A | 34582790 | No |
| PPP1R21 | 38356149 | No |
| PPP1R35 | 34582790 | No |
| PPP1R35 | 36598158 | Yes |
| PPP1R3F | 37531237 | Yes |
| PPP2R1A | 34582790 | No |
| PPP2R5C | 38258669 | Yes |
| PREX2 | 34582790 | No |
| PTCHD2 | 34582790 | No |
| PYCR2 | 34582790 | No |
| RAC3 | 35851598 | Yes |
| RAD21 | 34582790 | No |
| RANBP3L | 34582790 | No |
| RARB | 37092537 | Yes |
| RASGRF2 | 34582790 | No |
| RBM10 | 34582790 | No |
| RCOR3 | 34582790 | No |
| RNASEH2A | 34582790 | No |
| RNASEH2B | 34582790 | No |
| RNU4-2 | 38991538 | No |
| ROBO1 | 35227688 | Yes |
| ROBO3 | 38585811 | No |
| ROBO3 | 34582790 | No |
| RPA1 | 34582790 | No |
| RSPH4A | 39606420 | No |
| RSPH4A | 39606420 | No |
| RSPO4 | 34582790 | No |
| RTN2 | 34582790 | No |
| RXRA | 38990107 | No |
| SCN1A | 34582790 | No |
| SCN7A | 34582790 | No |

| Gene | PMID | Flag |
|------|------|------|
| *SEMA3F* | 38585811 | No |
| *SERAC1* | 34582790 | No |
| *SERPINB8* | 36622818 | No |
| *SETX* | 34582790 | No |
| *SF3B1* | 38258669 | Yes |
| *SHANK3* | 34582790 | No |
| *SHROOM4* | 34582790 | No |
| *SHROOOM3* | 39606420 | No |
| *SLC12A5* | 38585811 | No |
| *SLC18A2* | 34582790 | No |
| *SLC19A3* | 34582790 | No |
| *SLC25A45* | 34582790 | No |
| *SLC30A7* | 35751429 | No |
| *SLC30A7* | 35751429 | No |
| *SLC37A1* | 34582790 | No |
| *SLC39A10* | 34582790 | No |
| *SLC4A10* | 38054405 | Yes |
| *SLC5A7* | 39135055 | No |
| *SLC6A1* | 34582790 | No |
| *SLC7A1* | 34582790 | No |
| *SMARCA1* | 34582790 | No |
| *SMC3* | 38297832 | No |
| *SMPD1* | 34582790 | No |
| *SNAPC4* | 36965478 | Yes |
| *SNX14* | 34582790 | No |
| *SORCS2* | 34582790 | No |
| *SOX11* | 34582790 | No |
| *SPAST* | 36103453 | No |
| *SPR* | 34582790 | No |
| *SRD5A3* | 34582790 | No |
| *SRSF1* | 34582790 | No |
| *SSH3* | 34582790 | No |
| *SSTR3* | 34582790 | No |
| *STOML1* | 34582790 | No |
| *STON2* | 34582790 | No |
| STRTS locus | 38714869 | Yes |
| *STUB1* | 34582790 | No |
| *SUCLA2* | 34582790 | No |
| *SUFU* | 34675124 | No |
| *SYDE1* | 36622818 | No |
| *SYNE1* | 38716726 | Yes |
| *SYNGAP1* | 34582790 | No |
| *TALDO1* | 34677006 | No |
| *TBC1D23* | 34582790 | No |
| *TBX6* | 31888956 | Yes |
| *TCEAL1* | 36368327 | Yes |
| *TERF2* | 34582790 | No |
| *TGFBR2* | 38585811 | No |

| Gene | PMID | Novel |
|------|------|-------|
| *THOC6* | 34582790 | No |
| *TLK1* | 38868186 | Yes |
| *TLR7* | 35477763 | Yes |
| *TMOD1* | 34582790 | No |
| *TNRC6B* | 34582790 | No |
| *TOR1A* | 36757831 | No |
| *TPH2* | 34582790 | No |
| *TRAK1* | 34582790 | No |
| *TRAPPC4* | 34582790 | No |
| *TRIM66* | 34582790 | No |
| *TRIT1* | 34582790 | No |
| *TRMT1* | 34582790 | No |
| *TRMT2B* | 34582790 | No |
| *TTC12* | 39606420 | No |
| *TTLL11* | 34582790 | No |
| *TUBA1A* | 34582790 | No |
| *TUBB* | 38585811 | No |
| *TUBB4A* | 38585811 | No |
| *TUBB4A* | 34582790 | No |
| *TUBB6* | 34582790 | No |
| *TUBGCP2* | 34582790 | No |
| *UBE3C* | 36401616 | No |
| *UBR5* | 38258669 | Yes |
| *ULK2* | 34582790 | No |
| *UPS54* | 34582790 | No |
| *VANGL1* | 38669183 | Yes |
| *VANGL2* | 38669183 | Yes |
| *VPS28* | 34582790 | No |
| *VRK3* | 34582790 | No |
| *VSTM2L* | 34582790 | No |
| *WARS2* | 34582790 | No |
| *WDR45B* | 35322404 | No |
| *WDR62* | 34582790 | No |
| *WDR7* | 34582790 | No |
| *WDR73* | 34582790 | No |
| *WDR81* | 34582790 | No |
| *WDR83OS* | 34582790 | No |
| *WNT10B* | 36035248 | No |
| *WRN* | 35534204 | No |
| *WSB1* | 34582790 | No |
| *XRN1* | 38258669 | Yes |
| *YWHAG* | 38491959 | No |
| *ZBTB34* | 38258669 | Yes |
| *ZC4H2* | 34816580 | No |
| *ZFHX3* | 38412861 | Yes |
| *ZFHX4* | 39148819 | Yes |
| *ZNF462* | 38585811 | No |
| *ZRSR2* | 34582790 | No |